\documentclass[a4paper,12pt]{article}
\usepackage{amsmath,amssymb,amscd,mathtools,array,empheq}
\usepackage[utf8]{inputenc}
\usepackage[T1]{fontenc}
\usepackage{lmodern}
\usepackage[a4paper,tmargin=3truecm,bmargin=3truecm,hmargin=1.6truecm]{geometry}
\usepackage{graphicx}
\usepackage{color}
\usepackage{slashed}
\usepackage[dvipsnames]{xcolor}
\usepackage[english]{babel}
\usepackage{bm}
\usepackage[colorlinks=true, pdfstartview=FitV, linkcolor=blue, citecolor=blue, urlcolor=blue]{hyperref}
\usepackage{setspace}

\numberwithin{equation}{section}

\usepackage[shortlabels]{enumitem}

\usepackage{hyperref}
\hypersetup{colorlinks=true, urlcolor=blue, citecolor=blue, linktoc=page}

\usepackage[shadow,backgroundcolor=red!20]{todonotes}

\usepackage[normalem]{ulem} %% barrer texte + maintien \emph
\newcounter{mnotecount}[section]
\renewcommand{\themnotecount}{\thesection.\arabic{mnotecount}}
\newcommand{\mnote}[1]%
{\protect{\stepcounter{mnotecount}}${}^{\text{\footnotesize$\bullet$\themnotecount}}$%
\marginpar{\raggedright\tiny$\!\!\!\!\!\!\,\bullet$\themnotecount: #1}}

\usepackage{tikz}
\usetikzlibrary{shapes,arrows,arrows.meta,matrix,decorations.markings,calc,babel,quotes,angles}
\makeatletter \g@addto@macro\@floatboxreset\centering \makeatother

\newcommand\encircle[1]{%
  \tikz[baseline=(X.base)] 
    \node (X) [draw, shape=circle, inner sep=0] {\strut #1};}

\DeclareMathOperator{\const}{const}
\DeclareMathOperator{\Vect}{Vect}
\DeclareMathOperator{\Diff}{Diff}

\DeclareMathOperator{\diracf}{\slashed{D}^0}
\DeclareMathOperator{\dirac}{\slashed{D}}

\numberwithin{equation}{section}

\newcommand{\R}{\mathbb{R}}
\newcommand{\C}{\mathbb{C}}

\newcommand{\angles}[1]{\left\langle #1 \right\rangle}

\newcommand{\SO}{\mathrm{SO}}

\newcommand{\so}{\mathfrak{so}}
\newcommand{\volg}{\sqrt{|\rg|}\,}
\newcommand{\Ric}{\mathrm{Ric}}

\newcommand{\ie}{\textit{i.e.} }
\newcommand{\half}{\frac{1}{2}}

\newcommand{\wh}{\widehat}

\newcommand{\rg}{\mathrm{g}}

\newcommand{\cF}{\mathcal{F}}

\newcommand{\cL}{\mathcal{L}}
\newcommand{\cM}{\mathcal{M}}

\newcommand{\cP}{\mathcal{P}}

\newcommand{\cS}{\mathcal{S}}

\newcommand{\nvol}{|\mathrm{Vol}(\rg)|}
\newcommand{\bx}{\mathbf{x}}
\newcommand{\bvarpi}{\boldsymbol{\varpi}}
\newcommand{\bomega}{\boldsymbol{\omega}}
\newcommand{\bOmega}{\boldsymbol{\Omega}}
\newcommand{\bpartial}{\boldsymbol{\partial}}
\newcommand{\bbeta}{\boldsymbol{\beta}}
\newcommand{\bgamma}{\boldsymbol{\gamma}}
\newcommand{\bsigma}{\boldsymbol{\sigma}}
\newcommand{\bnabla}{\boldsymbol{\nabla}}
\newcommand{\bb}{\boldsymbol{b}}
\newcommand{\bc}{\boldsymbol{c}}
\newcommand{\bS}{\mathbf{S}}

\begin{document}

\title{On the L\'evy-Leblond-Newton equation
and its symmetries: a geometric view.
\\[5mm]}
 
\author{S. Lazzarini\footnote{mailto: lazzarini@cpt.univ-mrs.fr} \, and \, L. Marsot\footnote{mailto: marsot@cpt.univ-mrs.fr}\\[1.em]
{\normalsize Centre de Physique Théorique}\\
{\normalsize Aix Marseille Univ, Université de Toulon, CNRS, CPT, Marseille, France.}%
} %authors

\date{{\footnotesize (\today)}}

\maketitle

\bigskip

\begin{abstract}
The L\'evy-Leblond-Newton (LLN) equation for non-relativistic fermions with a gravitational self-interaction is reformulated within the framework of a Bargmann structure over a $(n+1)$-dimensional Newton-Cartan (NC) spacetime. The Schrödinger-Newton (SN) group introduced in~\cite{SN} as the maximal group of invariance of the SN equation, turns out to be also the group of conformal Bargmann automorphisms preserving the coupled L\'evy-Leblond and NC gravitational field equations. Within the Bargmann geometry a generalization of the LLN equation is provided as well. The canonical projective unitary representation of the SN group on 4-component spinors is also presented. In particular, when restricted to dilations, the value of the dynamical exponent $z=(n+2)/3$ is recovered as previously derived in \cite{SN} for the SN equation. Subsequently, conserved quantities associated to the (generalized) LLN equation are also exhibited.  
\end{abstract}

\vfill

\noindent{\bfseries Keywords:} L\'evy-Leblond fermions, Bargmann structure, dynamical exponent.%, @@@Cartan formalism ??


\smallskip
\noindent{\bfseries AMS classification scheme numbers:} 51P05,

\newpage

\tableofcontents

%%%%%%%%%%%%%%%%%%%%%%%%%%%%%%%%%
\section{Introduction}
\label{sec-Intro}
%%%%%%%%%%%%%%%%%%%%%%%%%%%%%%%%%

Penrose's original idea of ``gravitizing Quantum Mechanics'' instead of ``quantizing Gravitation'' \cite{Pen1,Pen2} is well illustrated by the example of the Schrödinger-Newton (SN) equation for a spinless massive particle, see {\it e.g.}~\cite{BGDB}, whose symmetries of the self-coupled system given by the (quantum) Schrödinger equation together with the (classical) Newton field equation have naturally been described in a Bargmann geometry~\cite{SN}. With the desire of incorporating the spin, the spin $1/2$ massive particle provides another example where Penrose's point of view can apply for incorporating a gravitational self-interaction. In a non-relativistic regime, L\'evy-Leblond fermions \cite{Leblond} turn out to be the natural candidates. Moreover, this type of particles has been receiving some attention in different contexts, see for instance \cite{DHP,Cariglia:2018rhw}.

%@article{Cariglia:2018rhw,
%      author         = "Cariglia, M. and Gibbons, G. W.",
%      title          = "{L\'evy-Leblond fermions on the wormhole}",
%      year           = "2018",
%      eprint         = "1806.05047",
%      archivePrefix  = "arXiv",
%      primaryClass   = "gr-qc",
%      SLACcitation   = "%%CITATION = ARXIV:1806.05047;%%"
%}
They ought to be also studied as slow neutrons in an ultra cold neutron beam for instance at the Institut Laue-Langevin (ILL-Grenoble, France) along the line as suggested by \cite{Neutrons}. This would yield another experimental test of Diòsi's first idea \cite{Dio} to give a role to gravity in quantum mechanics, whenever the Earth's gravitation field can be screened (as in free fall). Experiments like those as proposed in %by MAQRO
 \cite{Kaltenbaek:2012gt,Kaltenbaek:2015kha,MAQRO}, could also be supported by experiments at ILL in order to reveal a wave packet reduction process with a major change in the spreading of wave packets around and above a critical mass of a system composed by L\'evy-Leblond fermions. 
 
As will be recalled in the main text, L\'evy-Leblond derived a system (which carries his name) of two coupled 1st order partial differential equations, see \eqref{ll_eq}, which factorize the Schrödinger equation for bispinors. The price to pay is the doubling of the number of bispinors. The latter will become an important ingredient in the geometrization of the L\'evy-Leblond (LL) equation as subsequently performed in~\cite{Kunzle:1985bj}. 

\bigskip
At a more fundamental level, in this paper we shall study the symmetries of what we call the L\'evy-Leblond-Newton (LLN) equation as describing L\'evy-Leblond fermions coupled to Newton-Cartan (NC) geometry through their gravitational self-interaction. To some extent, one may consider the LLN equation as the ``square root'' of the SN equation. As such, it is rather natural to ask oneself whether the L\'evy-Leblond fermions can also be treated in the Bargmann framework. Previous indications in that direction were shown in \cite{Kunzle:1985bj,Duv4}. In particular, one may wonder which scale laws L\'evy-Leblond fermions are subject to, and which dynamical exponent in any spatial dimension characterizes them along the seminal idea given in~\cite{GG} for the SN equation. As a major result of the paper, the latter turns out to be the same as for the SN case, as computed in~\cite{SN}.

%Barut et al. \cite{Barut:1993nz}
%\bibitem{Barut:1993nz}
%  A.~O.~Barut, D.~J.~Moore and C.~Piron,
%  ``The Cartan formalism in field theory,''
%  Helv.\ Phys.\ Acta {\bf 66} (1993) 801
%   [Helv.\ Phys.\ Acta {\bf 66} (1993) 795].

In this respect, we shall mainly follow the line given in a previous work \cite{SN} in which most of the Bargmann study for the SN equation has been introduced. The reader will often be referred to the latter. The present paper is organized as follows. 
In Section \ref{sec-Prelim}, a quick review is made about the Bargmann geometry over a Newton-Cartan spacetime. Some delicacies are required in dealing with spinorial densities in order to have a correct geometrical description for the Dirac operator, the covariant derivative and the infinitesimal transformation (Lie derivative) of spinors.
Next, Section \ref{sec-LLN} is devoted to the L\'evy-Leblond-Newton coupled system along the line given in \cite{SN}. In particular, the generalized LLN equation is discussed in relation with gauge transformations.
Section \ref{sec-LLN-sym} treats the symmetries of the LLN equation collected in the SN group with in addition the corresponding spinorial representation. 
Explicit representations of this group will be given for spatially flat Bargmann structures. Of course, the corresponding projective unitary representation on LL spinors which is of importance at the quantum level is given. Also, conserved quantities of the LLN equation are exhibited.
Conclusions and some remarks are gathered in Section \ref{sec-conclusion}.

%%%%%%%%%%%%%%%%%%%%%%%%%%%%%%%%%%%%%%%%%%%%%%%%%%%%%%%%%%%%%%%%%%%%

\section{Preliminaries}
\label{sec-Prelim}

%%%%%%%%%%%%%%%%%%%%%%%%%%%%%%%%%%%%%%%%%%%%%%%%%%%%%%%%%%%%%%%%%%%%

\subsection{Bargmann structure and its link to Newton-Cartan}
\label{ss_barg}

It is usually convenient to write down non-relativistic systems, for example the Schr\"odinger equation, in the formalism of what is called a Bargmann structure \cite{Bargmann, BargD, Einsenhart}. This Lorentzian structure, which has one more dimension than the usual non relativistic $n+1$ Newton-Cartan spacetime, possesses geometrical tools which make the study of non-relativistic systems more geometrical and much easier to handle.

Let us recall that a Bargmann structure is  a principal bundle over spacetime, of one dimension higher. This structure is defined as a manifold $M$ endowed with a Lorentzian metric $\rg$ and a light-like vector field $\xi$, nowhere vanishing, with $\rg(\xi, \xi) = 0$. It is also equipped with the usual Levi-Civita connection, compatible with $g$ and $\xi$, such that $\nabla \rg = 0$ and $\nabla \xi = 0$. A Bargmann structure will then be denoted by the triple $(M, \rg, \xi)$.

\newcounter{w}
\newcounter{p}
\newcounter{h}
\setcounter{w}{9}
\setcounter{p}{1}
\setcounter{h}{7}

\tikzstyle{hidden} = [dashed,line width=1.1pt]
\tikzstyle{lesser} = [line width=1.2pt]
\tikzstyle{normal} = [line width=0.8pt]
\tikzstyle{normalh} = [dashed,line width=0.8pt]
\tikzstyle{arrow} = [line width=0.9pt, draw, -latex']
\tikzstyle{cone} = [line width=0.7pt]
\tikzstyle{labels} = [->]
\tikzstyle{carr} = [black!50!blue]
\tikzstyle{line} = [draw, -latex']
\tikzstyle{nc} = [black!50!red]

\tikzset{middlearrow/.style={
        decoration={markings,
            mark= at position #1 with {\arrow{{>}[scale=1.5]}} ,
        },
        postaction={decorate}
    }
}

\begin{figure}[ht]
\begin{tikzpicture}[line width=1.4pt,scale=0.85, every node/.style={transform shape}]
  % Top of barg
  \draw [lesser] (0,0) -- (0.45 * \value{w},\value{p}) -- (\value{w},0);
  \draw (\value{w},0) -- (0.55 * \value{w},-\value{p}) -- (0,0);

  % Vertical lines
  \draw [line width=1.4pt] (0,0) -- (0,-\value{h});
  \draw (0.55 * \value{w}, -\value{p}) -- (0.55 * \value{w}, -\value{p} - \value{h});
  \draw (\value{w}, 0) -- (\value{w}, -\value{h});
  \draw [hidden] (0.45 * \value{w},\value{p}) -- (0.45 * \value{w},\value{p} - \value{h});
  \draw [normal] (0.9 * \value{w}, - \value{p}) circle (0.6cm) node [scale=1.5]{$M$};

  % Bottom of barg
  \draw [hidden] (0, - \value{h}) -- (0.45 * \value{w},\value{p} - \value{h}) -- (\value{w}, - \value{h});
  \draw (\value{w}, - \value{h}) -- (0.55 * \value{w},-\value{p} - \value{h}) -- (0, - \value{h});
  
  % NC
  \draw [lesser,nc] (0, - 1.4 * \value{h}) -- (0.45 * \value{w},\value{p} - 1.4 * \value{h}) -- (\value{w}, - 1.4 * \value{h});
  \draw [nc] (\value{w}, - 1.4 * \value{h}) coordinate (nc1) -- node[pos=1,above,scale=1.5]{$\cM$} (0.55 * \value{w},-\value{p} - 1.4 * \value{h}) coordinate (nc2) -- (0, -1.4 * \value{h}) coordinate (nc3);
  \draw [normal] pic["",draw=black,-,angle eccentricity=1.2,angle radius=0.85cm] {angle=nc1--nc2--nc3};

  % Carroll
  \draw [normal,carr] (0.725 * \value{w}, 0.5*\value{p}) -- (0.275 * \value{w}, -0.5*\value{p}) coordinate (c1);
  \draw [normal,carr] (0.275 * \value{w}, -0.5*\value{p}) -- node[pos=0.92,right,scale=1.5]{$\widetilde{\Sigma}_t$} (0.275 * \value{w}, -1*\value{h} -0.5*\value{p});
  \draw [normalh,carr] (0.725 * \value{w}, 0.5*\value{p}) -- (0.725 * \value{w}, 0.5*\value{p} - \value{h}) coordinate (c3) -- (0.275 * \value{w}, -\value{h}-0.5*\value{p}) coordinate (c2);
  \draw [normal] pic["",draw=black,-,angle radius=1.25cm] {angle=c3--c2--c1};

  % Euclidean space
  \draw [normal, middlearrow={0.82},black!60!green] (0.725 * \value{w}, 0.5 * \value{p} - 1.4 * \value{h}) -- node [left,pos=0.32,scale=1.1] {$(x, t)\qquad$} node[scale=3.5,pos=1]{.} (0.05 * \value{w}, -1.4 * \value{h} - 1 * \value{p});
  
  % Axe temporel
  \draw [line] (-0.225 * \value{w}, -1.4 * \value{h} - 0.5* \value{p}) -- node [pos=1.18,below,scale=1.1] {$T \cong \R$ (time axis)} node[below, pos=0.58,scale=1.1]{$t = \const \qquad$} (0.325 * \value{w}, -1.5 * \value{p} - 1.4 * \value{h});
  
  % \xi
  \draw [arrow] (0.1 * \value{w}, -0.5 * \value{h}) -- node [near end, right,scale=1.1] {$\xi$} (0.1 * \value{w}, -0.38 * \value{h});
  \draw [arrow] (0.80 * \value{w}, -0.65 * \value{h}) -- node [near end, left, scale=1.1] {$\xi$} (0.80 * \value{w}, -0.53 * \value{h});
  \draw [arrow] (0.40 * \value{w}, -0.4 * \value{h}) -- node[pos=0,scale=3]{.} node [left,pos=0,scale=1.1]{$(x,t,s)$} node [near end, right,scale=1.1] {$\xi$} (0.40 * \value{w}, -0.28 * \value{h});

  % cône
  \draw [dashed,cone] (0.80 * \value{w}, -0.65 * \value{h}) -- (0.80 * \value{w}, -0.77 * \value{h});
  \draw [dashed,cone] (0.80 * \value{w}, -0.65 * \value{h}) -- (0.745 * \value{w}, -0.74 * \value{h});
  \draw [cone] (0.80 * \value{w}, -0.65 * \value{h}) -- (0.855 * \value{w}, -0.56 * \value{h});
  \draw [cone,rotate around={160:(0.827 * \value{w}, -0.55 * \value{h})}] (0.827 * \value{w}, -0.55 * \value{h}) ellipse (0.029 * \value{w} and 0.02 * \value{h});
  \draw [cone,dashed,rotate around={160:(0.773*\value{w}, -0.75 * \value{h})}] (0.773*\value{w}, -0.75 * \value{h}) ellipse (0.0295 * \value{w} and 0.02 * \value{h});
  
  % Labels  
  \node[draw, align=center,scale=1.1] at (1.3*\value{w}, \value{p}) (barg) {Extended\\Bargmann\\space-time-action\\$(M, \rg, \xi)$};
  \draw [labels] (barg) -- (0.83 * \value{w}, -0.1 * \value{p});
  
  \node[draw=black!50!blue, align=center,scale=1.1] at (-0.20 * \value{w}, \value{p}) (carr) {Carroll\\space-action\\$(\widetilde{\Sigma}_t, \Upsilon, \widetilde{\xi})$};
  \node[draw=none] at (0.4 * \value{w}, -0.5*\value{p}) (carr2) {};
  \draw [labels] (carr) to [out=0,in=100] (carr2);
  
  \node[draw=black!60!green, align=center,scale=1.1] at (1*\value{w}, -\value{h} - 1.5*\value{p}) (euclide) {Euclidean\\space\\$(\Sigma_t, h)$};
  \node[draw=none] at (0.55*\value{w}, -1*\value{h}-2.83*\value{p}) (euclide2) {};
  \draw [labels] (euclide) to [out=180,in=45] (euclide2);
  
  \node[draw=black!50!red, align=center,scale=1.1] at (1.1 * \value{w}, -\value{h} - 4.1 * \value{p}) (nc) {Newton-Cartan\\space-time\\$(\cM, h, \theta, \nabla^\cM)$};
  \node[draw=none] at (0.8 * \value{w}, -\value{h} - 2.95 * \value{p}) (nc2) {};
  \draw [labels] (nc) to [out=140,in=25] (nc2);
  
  \draw [normalh, middlearrow=0.4] (0.40 * \value{w}, -0.4 * \value{h}) -- (0.40 * \value{w}, -\value{h} - 0.75 * \value{p});
  \draw [normal, middlearrow=0.36] (0.40 * \value{w}, -\value{h} - 0.75 * \value{p}) -- node [left,pos=0.32,scale=1.5]{$\pi$} node [pos=1,scale=3]{.} (0.40 * \value{w}, -1.12 * \value{h} - 2.22 * \value{p});
\end{tikzpicture}
\caption{Visualization of a 1+2 dimensional Bargmann structure, and its link to Newton-Cartan and Carroll structures \cite{Marsot}.}
\label{f:barg}
\end{figure}

A characteristic of such structures is that the quotient $\cM = M/\R\xi$ is endowed with a Newton-Cartan structure \cite{Cartan,Kunzle,BargD}, or non-relativistic spacetime. In other words, the Newton-Cartan structure is a projection of the Bargmann structure along the direction given by the vector $\xi$. The projection yields a degenerate contravariant ``metric'' $h$, and a ``clock'' $\theta = dt$, generating the kernel of $h$, obtained from the projection of $\rg(\xi)$. Finally, the connection $\nabla^\cM$ on the Newton-Cartan structure is also given by the reprojection of the one living in Bargmann spacetime, and can be used to represent the gravitational field \cite{Cartan,Trautman}. Such a Newton-Cartan structure is denoted by the quadruple $(\cM, h, \theta, \nabla^\cM)$. Taking a section $t = \const$, called $\Sigma_t$, of this non relativistic spacetime naturally gives a Euclidean manifold $(\Sigma_t, h)$.

It is worthwhile to notice that another kind of structures can be recovered from Bargmann spacetime: those of Carroll \footnote{The name, coined by L\'evy-Leblond, refers to Lewis Carroll.} \cite{Leblond,Dautcourt,DGHBMS,DGH,DGHZ}. Each slice of constant time $\widetilde{\Sigma}_t$ in the Bargmann structure is endowed with a Carrollean structure. Because $\xi$ on Bargmann is light-like, the ``metric'' $\Upsilon$ induced on these Carroll space-actions is again degenerate because the induced vector field $\widetilde{\xi}$  generates the kernel of the metric. A Carroll structure is denoted by the triple $(\widetilde{\Sigma}_t, \Upsilon, \widetilde{\xi})$. All these geometries are sketched in Fig.\ref{f:barg}.

\medskip
Les us go back to Bargmann space. It can be equipped with coordinates $(x^1, \ldots, x^n, t, s)$, where $s$ has the dimension of an action (per mass). Locally, the metric $\rg$ and the vector field $\xi$ can be written as what is known as a Brinkmann metric \cite{Brink}:
\begin{equation}
\label{def_g_barg}
\rg = \rg_{{}_{\Sigma_t}} + dt \otimes \omega + \omega \otimes dt, \qquad \omega = \varpi_i(x, t) dx^i - U(x, t) dt + ds, \qquad \xi = \partial_s
\end{equation}
with $\rg_{{}_{\Sigma_t}} = \rg_{ij}(x, t) \, dx^i \otimes dx^j$ with $x = (x^1, \ldots, x^n)$ and where $\omega$ is a connection form on the principal $(\mathbb{R},+)$-bundle $\pi:M\to{}\cM$, with coefficients not depending on $s$. As said in \cite{SN}, the spacetime function~$U$ (for example, the profile of the gravitational wave whose wave-vector $\xi$ is null and parallel) is interpreted in the present context as the Newtonian gravitational potential on NC spacetime, $\cM$.
The functions $\varpi_i$ can be interpreted as some kind of Coriolis potential \cite{Ehlers,Costa15}. It can be thought of as a gravitational magnetic moment. They have physical dimension $[\varpi] = L T^{-1}$, and the Coriolis curvature $\Omega = d_{\Sigma_t} \varpi$ that will appear below in the Christoffel symbols \eqref{christo} has dimension $[\Omega] =  T^{-1}$. The Coriolis curvature $\Omega$ was shown to be relevant in various physical situations \cite{Costa15}, and it also appears in the Newtonian limit of the Taub-NUT spacetime \cite{Ehlers}\footnote{\label{fnnut}In this limit, one obtains the curvature $\Omega$ as $\Omega = \star d_{\R^3}\frac{a}{r}$, with $a$ the Taub-NUT parameter. We can recover $\varpi$ by analogy with a magnetic monopole. Indeed, solving for $\varpi$ is the same equation as solving for the potential vector in the case where the magnetic field is given by a magnetic monopole. One solution for the corresponding connection is therefore given by $\varpi_\mp = \frac{y dx - x dy}{r(z\pm r)}$ \cite{GockelerSchucker}.}.
While one could think that $\varpi$ could appear in the Kerr spacetime, this spacetime does not have a physically acceptable Newtonian limit, unless one wants to add the concept of negative mass \cite{Keres,Ehlers}.

\smallskip
For example, the simple flat case is given by the metric $\rg_0$ such that:
\begin{equation}
\label{barg_flat}
\rg_0 = \delta_{ij} dx^i \otimes dx^j + dt \otimes ds + ds \otimes dt\ .
\end{equation}
By projection \cite{BargD}, it induces the flat Newton-Cartan spacetime, given by $\cM=T\times\mathbb{R}^n$
with $(t,x^1,\ldots,x^n)$ as local coordinates, $h=\delta^{ij}\,\partial_i\otimes\partial_j$ and $\theta=dt$.

\smallskip
It is noteworthy to study what happens when the coordinate $s$ is transformed while preserving the fiber characterized by $\xi = \partial_s$. A general transformation of this kind is of the form
\begin{equation}
\label{trsf_s}
s \mapsto s + f(x,t),
\end{equation}
where $f$ is some function of $\cM$. Under this transformation, only the connection form $\omega$ from \eqref{def_g_barg} is modified in $\widehat{\omega}$ according to
\begin{equation}
\omega \mapsto \widehat{\omega} = \big(\varpi_i(x, t) - \partial_i f(x,t) \big) dx^i - \big(U(x, t) + \partial_t f(x,t)\big) dt + ds.
\end{equation}
Hence, in the general case, one is thus free to kill either \emph{one}  of the $n$ functions $\varpi_i$ or $U$ with such transformations.

In the particular case where $\varpi$ is exact, \ie $\varpi = d_{\Sigma_t}\vartheta$, with $\vartheta \in C^\infty(\cM,\R)$, one can turn off the terms $\varpi_i$ with such transformations~\eqref{trsf_s}\footnote{One could impose this form for $\varpi$ by postulating an additional field equation, due to Trautman \cite{Trautman}, namely~${R^{\mu\nu}}_{\lambda\rho} = 0$, and thus ignore $\varpi$ \cite{Ehlers}.}. Pushing this particular case further, if $\varpi = d_{\Sigma_t}\vartheta$ and $U = \partial_t \vartheta$, we can then turn off all gravitational and Coriolis potentials. 
We shall come back to this point in section~\ref{subsec-LLNonNC} in relation with gauge transformations.

\smallskip
Throughout the paper, $n = \dim \Sigma_t$ will be used as the spatial dimension, while $N = n + 2 = \dim M$ will be used for the dimension of the Bargmann space $M$.

%%%%%%%%%%%%%%%%%%%%%%%%%%%%%%%%%%%%%%%%%%%%%%%%%%%%%%%%%%%%%%%%%%%%

\subsection{Dirac operator and spinor densities}

Let $(M, \rg, \xi)$ be a N-dimensional Bargmann manifold, with $N = n + 2$. The Dirac operator was originally introduced by Dirac to describe relativistic quantum mechanics with spin. This operator, that we denote $\diracf(\rg) = \gamma^\mu \nabla_\mu$ acts on spinors  $\psi \in \cS(M) = L^2(M)\otimes \C^k$, with $N = 2 k$ if $N$ is even, and $N = 2 k + 1$ if $N$ is odd, and with $(\gamma^\mu)$ the set of gamma matrices belonging to the Clifford algebra associated to the Bargmann space, such that $\gamma^\mu \gamma^\nu + \gamma^\nu \gamma^\mu = - 2 \rg^{\mu\nu}$ \footnote{Mind the sign convention.\label{sign-footnote}}.

This Dirac operator $\diracf(\rg)$ transforms non trivially under conformal dilations (or Weyl rescalings) of the metric $\rg \rightarrow \wh \rg = \lambda \rg$, with $\lambda \in C^\infty(M, \R_+^*)$. Yet, we would like to have a conformally invariant operator. We will write $\dirac(\rg)$ an operator satisfying
\begin{equation}
\label{inv_conf_dirac}
\dirac(\wh \rg) = \dirac(\rg)\ .
\end{equation}

It turns out that such a conformally invariant operator $\dirac(\rg)$ can be constructed from the usual Dirac operator $\diracf(\rg)$ in the following way grounded on geometry, see \cite{MRS}
\begin{equation}
\label{rel_dirac}
\dirac(\rg) = \nvol^{\frac{N+1}{2N}} \circ \diracf(\rg) \circ \nvol^{- \frac{N-1}{2N}},
\end{equation}
where $\nvol$ is the canonical volume element of $M$. These volume forms cancel out the non trivial dilation terms coming from $\diracf(\rg)$. However, definition \eqref{rel_dirac} has a cost, now our operator $\dirac(\rg)$ does not act on spinors anymore, it rather acts on spinor densities which will be denoted by $\Psi \in \cS_w(M) = \cS(M) \otimes \cF_w(M)$, where $\cF_w(M)$ stands for the space of densities of weight $w$. This means that locally, spinor densities are written as
\begin{equation}
\label{spinor_densities}
\Psi = \psi \, \nvol^w,
\end{equation}
where $\psi \in \cS(M)$. Notice that usual spinors $\psi$ are merely 0-densities.

\tikzstyle{hidden} = [dashed,line width=1.1pt]
\tikzstyle{lesser} = [line width=1.2pt]
\tikzstyle{normal} = [line width=0.8pt]
\tikzstyle{normalh} = [dashed,line width=0.8pt]
\tikzstyle{arrow} = [line width=0.9pt, draw, -latex']
\tikzstyle{cone} = [line width=0.7pt]
\tikzstyle{labels} = [->]
\tikzstyle{carr} = [black!50!blue]
\tikzstyle{line} = [draw, -latex']
\tikzstyle{nc} = [black!50!red]

\tikzset{middlearrow/.style={
        decoration={markings,
            mark= at position #1 with {\arrow{{>}[scale=1.5]}} ,
        },
        postaction={decorate}
    }
}
\begin{figure}[ht]
\begin{tikzpicture}[line width=1.4pt,scale=1.1, every node/.style={transform shape}]
  \node [draw=white,align=center] at (0, 0) (fl) {$\cS_{w=\frac{N-1}{2N}}(M)$};
  \node [draw=white,align=center] at (5, 0) (fm) {$\cS_{w'=\frac{N+1}{2N}}(M)$};
  \node [draw=white,align=center] at (0, -3) (cinf1) {$\cS(M)$};
  \node [draw=white,align=center] at (5, -3) (cinf2) {$\cS(M)$};

  \draw [arrow] (fl) to node [pos=0.5,above] {$\slashed{D}(\rg)$} (fm);
  \draw [arrow] (cinf1) to node [pos=0.5,left] {$\nvol^w$} (fl);
  \draw [arrow] (cinf2) to node [pos=0.5,right] {$\nvol^{w'}$} (fm);
  \draw [arrow] (cinf1) to node [pos=0.5,below] {$\slashed{D}^0(\rg)$} (cinf2);
\end{tikzpicture}
\caption{Relations between the spaces of spinors, spinor densities, and both Dirac operators}
\label{f:link_diracs}
\end{figure}

With the definitions \eqref{rel_dirac} and \eqref{spinor_densities}, and the fact that the spinor densities are of weight $w = \frac{N-1}{2N}$ here, it is easy to see the action of the conformally invariant Dirac operator on spinor densities,
\begin{equation}
\label{dirac_spinor}
\dirac(\rg) \Psi = \left(\diracf(\rg) \psi\right) \nvol^\frac{N+1}{2N},
\end{equation}
meaning that the operator $\dirac(\rg)$ sends $\frac{N-1}{2N}$-densities to $\frac{N+1}{2N}$-densities, while the action on the spinorial part is just like the usual Dirac operator, as summed up in the diagram in Fig.\ref{f:link_diracs}.

\medskip
Let us now consider the action of covariant derivatives on spinor densities. Since the connection used here is the usual Levi-Civita connection, it is compatible with the metric and thus the covariant derivative only sees the spinorial part, and not the volume. Its action on a spinor $\psi \in \cS(M)$ is defined as \cite{Kosmann,Souriau64}
\begin{equation}
\label{der_spinors}
\nabla_X \psi = X^\mu \partial_\mu \psi - \frac{1}{8} X^\mu \left[\gamma^\rho, \partial_\mu \gamma_\rho - \Gamma^\sigma_{\mu\rho} \gamma_\sigma\right] \psi,
\end{equation}
with $\gamma_\mu = \rg_{\mu\nu} \gamma^\nu$, and is such that $\nabla \gamma = 0$.

\medskip
The action of a Lie derivative along a vector field $X$ on a spinor can also be defined \cite{Kosmann},
\begin{equation}
\label{def_lie_spinor}
L_X \psi = \nabla_X \psi - \frac{1}{4} \gamma^\mu \gamma^\nu \nabla_{[\mu} X_{\nu]} \, \psi.
\end{equation}
Recall also the action of a Lie derivative on the volume element,\begin{equation}
\label{lie_volume}
L_X \nvol = \left(\nabla_\mu X^\mu \right) \nvol
\end{equation}
so that, in order to obtain the action of a Lie derivative on a spinor density, one must combine \eqref{def_lie_spinor} and \eqref{lie_volume} with \eqref{spinor_densities}, to get
\begin{equation}
\label{lie_spinor_density}
L_X \Psi = \left( \nabla_X \psi \right)\nvol^\frac{N-1}{2N} - \frac{1}{4} \gamma^\mu \gamma^\nu \nabla_{[\mu} X_{\nu]} \Psi + \frac{N-1}{2N} \left(\nabla_\mu X^\mu\right) \Psi \ .
\end{equation}

Notice for later use, that in the case of a Lie derivative along $\xi$, the covariantly constant null vector field entering in the definition of a Bargmann structure, we get that $L_\xi \Psi = \left(L_\xi \psi \right) \nvol^\frac{N-1}{2N}$.

In conclusion, working with spinor densities is almost transparent, most operators used here act on spinor densities just like they do on spinors. We still have to be careful when considering dilations and conformal transformations, as the densities will play an important role there.

%%%%%%%%%%%%%%%%%%%%%%%%%%%%%%%%%%%%%%%%%%%%%%%%%%%%%%%%%%%%%%%%%%%%
\section{The L\'evy-Leblond--Newton system}
\label{sec-LLN}

It is recalled that the L\'evy-Leblond equation \cite{LL} is an equation describing non relativistic fermions in 3-dimensional space. While one could have worked with the Schr\"odinger-Pauli equation, L\'evy-Leblond showed that the Schr\"odinger equation can be factorized into a system of first order partial differential equations, in analogy with the derivation of the Dirac equation from the Klein-Gordon's one. The free L\'evy-Leblond system of PDE's is given by
\begin{equation}
\label{ll_eq}
\left\lbrace\begin{array}{l}
\hbar \, \sigma(\bpartial) \, \varphi + 2 \, m \, \chi = 0 \\[1ex]
i \, \hbar \, \partial_t \varphi - \hbar \, \sigma(\bpartial) \, \chi = 0
\end{array}\right.
\end{equation}
with two bispinors $\varphi$ and $\chi$, where $\sigma$ denotes the set of the three Pauli matrices, and\footnote{Throughout the paper,   the notation $\sigma(\boldsymbol{a}) = \sigma_k a^k = \vec{\sigma}\!\cdot\!\vec{a}$, where bold letters $\boldsymbol{a}=\vec{a}$ for vectors in $\mathbb{R}^3$, will be used. A slight abuse of notation yields $\boldsymbol{\sigma}$ as well to compactly denote the three Pauli matrices.} 
$\sigma(\bpartial) = \sigma^i \partial_i$. It is worthwhile to notice that the second bispinor $\chi$ is non-dynamical, unlike in the Dirac equation. This is to be expected since the Schr\"odinger equation is of first order in time. The two bispinors fit into a 4-spinor $\psi = \left(\begin{array}{c}\varphi \\ \chi \end{array}\right)$.

The L\'evy-Leblond--Newton equation, or LLN for short, is when we add a gravitational potential in the L\'evy-Leblond equation, whose source is the probability density of the 4-spinor $\psi$. This is in the same spirit of the Schr\"odinger-Newton equation \cite{Dio,SN}, which is the Schr\"odinger equation with a gravitational potential whose source is the probability density of the wavefunction. The system thus becomes,
\begin{equation}
\label{lln}
\left\lbrace\begin{array}{l}
\hbar \, \sigma(\bpartial) \, \varphi + 2 \, m \, \chi = 0 \\[1ex]
i \, \hbar \, \partial_t \varphi - m U \varphi - \hbar \, \sigma(\bpartial) \, \chi = 0
\end{array}\right.
\end{equation}
together with the Poisson equation for the potential $U$ and mass density $\rho$
\begin{equation}
\label{poisson}
\Delta U(x, t) = 4 \pi G \rho, \qquad \rho = m \varphi^\dagger \varphi\, .
\end{equation}

While one could work with this system directly, we will see that writing the LLN equation in the formalism of a Bargmann structure will make the symmetries apparent and the general study of this system more transparent.

%%%%%%%%%%%%%%%%%%%%%%%%%%%%%%%%%%%%%%%%%%%%%%%%%%%%
\subsection{Lifting LLN on the Bargmann space}

Motivated by the previous considerations, let us call the L\'evy-Leblond--Newton system on Bargmann the set of coupled equations\footnote{Notice also that the covariant form of the LL equation was first provided in~\cite{Kunzle:1985bj}.}
\begin{subequations} \label{lln_barg}
\begin{empheq}[left=\empheqlbrace]{align}
& \gamma^\mu \gamma^\nu + \gamma^\nu \gamma^\mu = - 2 \rg^{\mu\nu} \label{clialg} \\[1ex]
& \dirac \Psi = 0 \label{diracmnul} \\[1ex]
& L_\xi \Psi = i \frac{m}{\hbar}\, \Psi \label{equivariance} \\[1ex]
& \Ric(\rg) = 4 \pi G \rho \, \theta \otimes \theta \label{poissonbarg} \\[1ex]
& \rho = m \overline\Psi^\sharp \gamma(\xi) \Psi^\sharp \label{density}
\end{empheq}
\end{subequations}
with $\Psi$ a spinor $\frac{N-1}{2N}$-density, that we can locally decompose as $\Psi = \psi \, \nvol^\frac{N-1}{2N}$, where here $\psi = \left(\begin{array}{c}\varphi \\ \chi \end{array}\right)$, with $\varphi$ and $\chi$ two bispinors. Then, $m$ is a mass, and $\overline \Psi = \Psi^\dagger G$, with $G$ such that $\overline \gamma_\mu = G^{-1} \gamma^\dagger_\mu G = \gamma_\mu$, and $G^\dagger = G$.
%Let us comment some of these equations.
Some comments are in order. 

\smallskip
The reader's attention is drawn to %Mind 
the sign in the relation \eqref{clialg} defining the Clifford algebra, (see footnote~\ref{sign-footnote}). This comes from the signature of the metric, chosen to be $(+, \ldots, +, -)$ on the Bargmann space, so that we recover a positive metric when projecting onto the non-relativistic Newton--Cartan spacetime. Also, while a seemingly arbitrary dimension $N$, or $n$, appears in the relations, the reader must keep in mind\footnote{We shall generically work in space dimension $n$, going back to $n=3$ when required.}
that this work is focused on the $n = 3$ case, namely $N=5$.

In the probability density definition \eqref{density}, the notation $\Psi^\sharp$ corresponds to a \emph{normalized} spinor which is defined as
\begin{equation}
\label{norm_psi}
\Psi^\sharp = \frac{\Psi}{||\Psi||_\rg} \qquad \& \qquad ||\Psi||^2_\rg = \int_{\Sigma_t} \overline \psi \gamma(\xi) \psi \, \mathrm{vol}( h)
\end{equation}
so that $||\Psi^\sharp||^2_\rg = 1$. 
Here $\mathrm{vol}(h)$ stands for the canonical volume form of $\Sigma_t$,\footnote{This volume form can be defined intrinsically. Indeed, call $\eta=\rg^{-1}(\omega)$ the vector field associated with the connection form $\omega$ given by \eqref{def_g_barg}; one checks that $\eta$ is null and $\omega$-horizontal. Then $\mathrm{vol}(\rg)(\xi,\eta)$ flows down to NC spacetime, $\cM$; once pulled-back to $\Sigma_t$, it canonically defines the volume $n$-form $\mathrm{vol}(h)$. 
The latter admits the following local expression, namely $\mathrm{vol}(h)=\sqrt{\det(\rg_{ij}(x,t))}\,dx^1\wedge\cdots\wedge dx^n$, where $h=h^{ij}(x,t)\,\partial/\partial{x^i}\otimes\partial/\partial{x^j}$ and $(h^{ij})=(\rg_{ij})^{-1}$.
}
and $\psi(\bx,t)=\psi_t(\bx)$ --- with $\psi_t\in{}L^2\left(\Sigma_t,\mathrm{vol}(h)\right)\otimes \mathbb{C}^k$ --- is as in \eqref{spinor_densities}.

\medskip
The intent to write this system on a Bargmann space and not directly on the usual non relativistic space time, Newton--Cartan, is that on Bargmann the system is written in a completely covariant and geometrical way, which makes it easier to compute its symmetries. Note that on Bargmann, this system is written with a Dirac equation for a null mass~\eqref{diracmnul}.

%%%%%%%%%%%%%%%%%%%%%%%%%%%%%%%%%%%%%%%%%%%%%%%%%%%%%%%%%%%%%%%%%%%%
\subsection{Recovering the LLN system on Newton--Cartan \label{subsec-LLNonNC}}

The L\'evy-Leblond equation \eqref{ll_eq}, as originally written \cite{LL}, was on flat space of dimension $n = 3$. In order to recover the LLN equation from the system \eqref{lln_barg}, we first put ourselves in this case, with a spatially flat metric on Bargmann space,
\begin{equation}
\label{barg_metric}
\rg = \Vert d\bx\Vert^2 + 2 \, dt \, ds - 2 \,U(\bx, t) dt^2 + 2 \,\bvarpi(\bx, t)\!\cdot\! d\bx \, dt
\end{equation}
with $\bx\in\mathbb{R}^3$, $U(\bx, t)$ a scalar potential, and $\bvarpi(\bx, t)$ a covariant Coriolis vector potential.

We are now going to see what each of the relations in the system \eqref{lln_barg} becomes when we specify the metric to \eqref{barg_metric}.

\paragraph{Clifford algebra}
In order to satisfy the Clifford algebra \eqref{clialg} for the Bargmann metric \eqref{barg_metric} whose matrix reads
\[
\rg = (\rg_{\mu\nu}) = \begin{pmatrix}
I_3 &  \bvarpi & \boldsymbol{0} \\
\bvarpi^{\text{T}} & -2U & 1\\
\boldsymbol{0}^{\text{T}} & 1 & 0
\end{pmatrix}, \qquad
\text{and} \quad 
\rg^{-1} = (\rg^{\mu\nu}) = \begin{pmatrix}
I_3 & \boldsymbol{0} & -\bvarpi \\
\boldsymbol{0}^{\text{T}} & 0 & 1\\
- \bvarpi^{\text{T}} & 1 & 2U+ \bvarpi^2
\end{pmatrix},
\]
the set of gamma matrices is computed to be 
\begin{equation}
\gamma^t = \left(\begin{array}{cc}
0 & 0 \\
1 & 0
\end{array}\right), \qquad
\gamma^j = \left(\begin{array}{cc}
- i \sigma^j & 0 \\
0 & i \sigma^j
\end{array}\right), \qquad
\gamma^s = \left(\begin{array}{cc}
i \sigma(\bvarpi) & -2 \\
U & -i \sigma(\bvarpi)
\end{array}\right)
\end{equation}
where the $\sigma^j$ are the Pauli matrices, $U = U(\bx, t)$, $\bvarpi = \bvarpi(\bx, t)$, and $\sigma(\bvarpi) = \sigma^i \varpi_i$. Note that since the metric is spatially flat, we have $\varpi^i = \varpi_i$, and likewise for $\boldsymbol{\sigma}$. We have $\gamma_\mu = \rg_{\mu\nu} \gamma^\nu$, which becomes,
\begin{equation}
\gamma_t = \left(\begin{array}{cc}
0 & -2 \\
- U & 0
\end{array}\right), \qquad
\gamma_j = \left(\begin{array}{cc}
- i \sigma_j & 0 \\
\varpi_j & i \sigma_j
\end{array}\right), \qquad
\gamma_s = \left(\begin{array}{cc}
0 & 0 \\
1 & 0
\end{array}\right),
\end{equation}
such that we also have $\gamma_\mu \gamma_\nu + \gamma_\nu \gamma_\mu = - 2 \rg_{\mu\nu}$.

\paragraph{Equivariance relation}
To compute the equivariance relation \eqref{equivariance}, we need the Christoffel symbols associated to the Bargmann metric \eqref{barg_metric}. The non-zero ones are, (see \cite{SN}):
\begin{align} \label{christo}
\Gamma^i_{tt} &= \partial_i U + \partial_t \varpi_i; \quad \Gamma^s_{it} = - \partial_i U - \half \Omega_{ij} \varpi^j; \notag\\[-3mm]
\\[-3mm]
\Gamma^s_{tt} &= - \partial_t U - \varpi^i(\partial_i U + \partial_t \varpi_i); \quad \Gamma^s_{ij} = \partial_{(i}\varpi_{j)}; \quad \Gamma^i_{jt} = - \half \Omega_{ij} \notag
\end{align}
with $\Omega = d_{\Sigma_t} \bvarpi$ the Coriolis curvature.
% \red{where $d$ is the differential on the Euclidean space~$\Sigma_t\simeq\mathbb{R}^n$}.

\smallskip
Since $\xi(U) = 0$, and $\xi^\mu \Gamma^\rho_{\mu\nu} = 0$, applying the definition of the Lie derivative on a spinor density \eqref{lie_spinor_density} gives $L_\xi \Psi = \partial_\xi \Psi = i \frac{m}{\hbar} \Psi$, exactly like in the free case where the potential $U$ and the Coriolis vector potential $\bvarpi$ vanish.

Note that this equivariance relation together with the density character of the spinors \eqref{spinor_densities} imply the following decomposition of $\Psi$,
\begin{equation}
\label{decomp_psi}
\Psi(\bx, t, s) = e^\frac{ims}{\hbar} \, \psi(\bx, t) \, \nvol^\frac{n+1}{2(n+2)}.
\end{equation}

\paragraph{Poisson equation}
From the metric  \eqref{barg_metric} used here, the Ricci tensor gives constraints on $U$ and $\omega$, so that the gravitation equation \eqref{poissonbarg} takes the form
\begin{equation}
\delta \Omega = 0 \qquad \& \qquad \Delta_{\R^n} U + \frac{\partial}{\partial t} \delta \bvarpi + \half \Vert\Omega\Vert^2 = 4 \pi G \rho
\end{equation}
with $\delta$ the codifferential acting on differential forms on the Euclidean space~$\Sigma_t\simeq\mathbb{R}^n$ and $\Vert\Omega\Vert^2 = \half \delta^{ik}\delta^{jl} \Omega_{ij}\Omega_{kl}$. Note that in the case $\bvarpi = 0$, we recover the usual Poisson equation \eqref{poisson}.\footnote{The non relativistic limit of Taub-NUT spacetimes, see footnote \ref{fnnut}, yields $\delta \Omega = 0$, $\delta \varpi = 0$, and $\Vert\Omega\Vert^2~=~4 a^2/r^4$.}

The density \eqref{density} is such that $\rho = m \overline\Psi \gamma(\xi) \Psi = m \Psi^\dagger G \gamma_s \Psi = m \varphi^\dagger \varphi$, with $G = \left(\begin{array}{cc} 0 & 1\\1 & 0
\end{array}\right)$. Note that the probability density only involves the first bispinor $\varphi$, as was remarked by L\'evy-Leblond in~\cite{LL}. This is not a problem, as we will see later on.

\paragraph{The massless Dirac equation}
We are now left with the massless Dirac equation on Bargmann \eqref{diracmnul}, $\dirac(\rg) \Psi = 0$.

The second term in the covariant derivative of spinors \eqref{der_spinors} can be split into two parts: $\gamma^\mu \left[\gamma^\rho, \partial_\mu \gamma_\rho\right]$ and $- \gamma^\mu \left[\gamma^\rho, \Gamma^\sigma_{\mu\rho} \gamma_\sigma\right]$. We have for the former
\begin{equation}
\label{eq-former}
\gamma^\mu \left[\gamma^\rho, \partial_\mu \gamma_\rho\right] = - 2 \sigma^i \sigma^j \partial_i \varpi_j \left(\begin{array}{cc}
0 & 0 \\
1 & 0
\end{array}\right),
\end{equation}
while the latter becomes
\begin{align}
- \left[\gamma^\rho, \Gamma^\sigma_{\mu\rho} \gamma_\sigma\right] & = \Gamma^i_{\mu t} [\gamma_i, \gamma^t] + \Gamma^s_{\mu t} [\gamma_s, \gamma^t] + \Gamma^s_{\mu i} [\gamma_s, \gamma^i] + \Gamma^i_{\mu j} [\gamma_i, \gamma^j] \notag\\[1mm]
& = 2 \, i \, \sigma^j \, \left(\Gamma^j_{\mu t} - \Gamma^s_{\mu j}\right) \left(\begin{array}{cc}
0 & 0 \\
1 & 0
\end{array}\right) - 2 \, i \, \Gamma^j_{t k} \left(\begin{array}{cc}
\epsilon_{j k l} \sigma^l & 0 \\
\sigma^k \varpi^j & \epsilon_{j k l} \sigma^l
\end{array}\right),
\end{align}
with $\epsilon_{j k l}$ the fully skewsymmetric Levi-Civita tensor, and $\epsilon_{1 2 3} = 1$. The non zero components are for $\mu = t$ and $\mu = j$; they read
\begin{align}
- \left[\gamma^\rho, \Gamma^\sigma_{t\rho} \gamma_\sigma\right] & = 2 \,i\, \sigma^k \left(2 \partial_k U + \partial_t \varpi_k\right) \left(\begin{array}{cc}0 & 0\\1 & 0\end{array}\right) + 2 \, i \, \epsilon_{k l m} \sigma^m \partial_k \varpi_l \left(\begin{array}{cc}1 & 0 \\ 0 & 1\end{array}\right),\\
- \left[\gamma^\rho, \Gamma^\sigma_{j\rho} \gamma_\sigma\right] & = - 2 \, i \, \sigma^k \partial_k \varpi_j \left(\begin{array}{cc}0 & 0\\1 & 0\end{array}\right).
\end{align}
Upon contracting with $\gamma^\mu$, we get,
\begin{equation}
- \gamma^\mu \left[\gamma^\rho, \Gamma^\sigma_{\mu\rho} \gamma_\sigma\right] = 2 \, \delta^i_j \, \partial_i \varpi^j \left(\begin{array}{cc}
0 & 0 \\
1 & 0
\end{array}\right),
\end{equation}
which combined with \eqref{eq-former} yields,
\begin{equation}
\gamma^\mu \left[\gamma^\rho, \partial_\mu \gamma_\rho - \Gamma^\sigma_{\mu\rho} \gamma_\sigma\right] = - 2 \, i \, \sigma\left(\bpartial\times\bvarpi\right) \left(\begin{array}{cc}
0 & 0 \\
1 & 0
\end{array}\right).
\end{equation}
The massless Dirac equation \eqref{diracmnul} on Bargmann can thus be developed as
\begin{align}
\left[ \left(\begin{array}{cc}
0 & 0 \\
\partial_t & 0
\end{array}\right)
\right. & +
\left(\begin{array}{cc}
- i \sigma(\bpartial) & 0 \\
0 & i \sigma(\bpartial)
\end{array}\right)  \notag \\[3mm]
& \left. +\,
\frac{i \, m}{\hbar} \left(\begin{array}{cc}
i \sigma(\bvarpi) & -2 \\
U & -i \sigma(\bvarpi)
\end{array}\right)
+
\frac{1}{4}\left(\begin{array}{cc}
0 & 0 \\
i \sigma(\bpartial\times\bvarpi) & 0
\end{array}\right)
\right]
\left(\begin{array}{c}
\varphi \\
\chi
\end{array}\right) = 0,
\end{align}
which generalizes the original LL equation since it equivalently reads in bispinor components as
\begin{equation}
\label{generalized_lln_eq}
\left\lbrace\begin{array}{l}
\hbar \, \sigma(\bpartial) \, \varphi + 2 \, m \, \chi - i \, m \, \sigma(\bvarpi)\, \varphi = 0 \\[1ex]
i \, \hbar \, \partial_t \varphi - m U \varphi - \hbar \, \sigma(\bpartial) \, \chi + i \, m \, \sigma(\bvarpi) \, \chi - \frac{1}{4} \hbar \, \sigma(\bpartial\times\bvarpi) \, \varphi = 0.
\end{array}\right.
\end{equation}
The first equation can be recast to show the 1st order relation between the two bispinors,
\begin{equation}
\label{rel_two_bispinors}
\chi = -\frac{\hbar}{2m} \sigma(\bpartial) \, \varphi + \frac{i}{2} \sigma(\bvarpi) \, \varphi \, ,
\end{equation}
and gives us the opportunity to write the system \eqref{generalized_lln_eq} solely in terms of the principal bispinor $\varphi$. This is the reason why writing the probability density  only in terms of $\varphi$ is not a problem, the second bispinor is somewhat redundant in the LL model. We thus recover a second order differential equation, akin to the Schr\"odinger equation, for a bispinor $\varphi$ with a gravitational potential $U$, and the Coriolis (co)vector potential $\bvarpi$,
\begin{equation} \label{eq-H}
\left(-\frac{\hbar^2}{2m} \Delta + \frac{i \hbar}{2}\left[\sigma(\bpartial)\circ\sigma(\bvarpi) + \sigma(\bvarpi)\circ\sigma(\bpartial)\right] + m \!\left(U + \frac{\Vert\bvarpi\Vert^2}{2}\right) + \frac{1}{4} \hbar \sigma(\bpartial\times\bvarpi) \right) \varphi = i\hbar \, \partial_t\, \varphi.
\end{equation}
It is worthwhile to notice at this stage that the (self-adjoint) Hamiltonian in the l.h.s. of \eqref{eq-H} fulfills the most general form dictated by the Galilean relativity principle as stated in \cite{Jauch,Piron} and refreshed in a modern language in \cite[§ 8.4 Galilean invariance]{LeBellac}. This principle provides a way to justify the minimal coupling form through the strong link between translation in momentum and the action of Galilean boosts. 
According to \cite{Piron} the most general form for a Hamiltonian acting on a bispinor is thus given by 
\[
H = \frac{1}{2m} \big( \mathbf{P}I_2 - \mathbf{A}_\mu(\bx,t) \sigma^\mu \big)^2 + V_\mu(\bx,t) \sigma^\mu
\]
where $\mathbf{P}$ is the momentum operator, for $\mu=0,1,2,3,4$, $\mathbf{A}_\mu$ gives four vector fields, $V_\mu$ stands for four scalar fields and $(\sigma^\mu) = (I_2,\bsigma)$ is a basis for $2\times 2$ complex matrices. After some algebra, a direct comparison yields (dropping the unit matrix) the equivalent expression\footnote{The Hamiltonian occurring in the generalized SN equation \cite[Eq.(3.9)]{SN} is readily seen to be recast into the canonical form as
$H =  \frac{1}{2m} \big( \mathbf{P} - m \bvarpi \big)^2 + m U$.} for the Hamiltonian obtained in \eqref{eq-H}\footnote{Since Galilean boosts form an abelian subgroup of the SN group, such a canonical form for the Hamiltonian was expected.}
\begin{equation}
\label{hamiltonien}
H = \frac{1}{2m} \big( \mathbf{P} - m \bvarpi \big)^2 + m U - \frac{1}{4}\, \hbar \,  \sigma(\bpartial\times\bvarpi)
\end{equation}
for $\mathbf{A}_0 =  m \bvarpi$, $\mathbf{A}_k \equiv 0$, $V_0 = m U$ and 
$\boldsymbol{V} = (V_1,V_2,V_3) = - \frac{1}{4}\, \hbar\, (\bpartial\times\bvarpi)$. The last term is reminiscent of the Pauli coupling term $\sigma(\boldsymbol{B})$ for spin $1/2$. Note that the (pseudo) vector $\bOmega = \bpartial\times\bvarpi$ is linked to the curvature 2-form $\Omega$ by $\bOmega = \star \Omega$. It remains to interpret the coupling upon setting $\mathbf{S} = \hbar\,\boldsymbol{\sigma}/2$ for the spin operator
\[
- \frac{1}{4}\, \hbar \,  \sigma(\bpartial\times\bvarpi) = -\half \, \bS\!\cdot\! \bOmega
\]
where $\bOmega = \bpartial\times\bvarpi$ is very similar to $\boldsymbol{B} = \bpartial\times \boldsymbol{A}$ for $\boldsymbol{A}$ the usual Maxwell vector potential. 

In order to complete the analogy with electromagnetism, we can look at spin precession due to this Coriolis term. Computing the usual time evolution of the operator through $d\bS/dt = \frac{i}{\hbar} \left[H, \bS\right]$ and the Hamiltonian \eqref{hamiltonien}, we obtain,
\begin{equation}
\frac{d\bS}{dt} = \half\, \bS \times \bOmega,
\end{equation}
in accordance with \cite{Costa15}.

\smallskip
On the other hand, thanks to the canonical form of the Hamiltonian $H$ given in \eqref{hamiltonien}, it is well-known that a $U(1)$-gauge transformation of the wave-function corresponds to a gauge transformation of the potentials, see {\it e.g.} \cite[§13-5]{Jauch}.
In light of these observations, one may wonder whether by a phase change on the bispinor $\varphi(\bx,t) \mapsto (\Theta\varphi)(\bx,t) = e^{\frac{im}{\hbar} \vartheta (\bx,t)} \varphi(\bx,t)$ the Coriolis potential could be put to zero. Mimicking \cite[§13-5]{Jauch}, for $\varphi$ subject to the Schrödinger equation $ i\hbar \, \partial_t\, \varphi= H\varphi$, one gets
\begin{align*}
&\Theta \mathbf{P} \Theta^{-1} = \mathbf{P} - m \bpartial \vartheta \quad\Rightarrow \quad \Theta \big( \mathbf{P} - m \bvarpi \big)^2 \Theta^{-1} = \big( \mathbf{P} - m (\bvarpi + \bpartial \vartheta) \big)^2 \\
& H' = \Theta H \Theta^{-1} + i\hbar (\partial_t \Theta) \Theta^{-1} = \frac{1}{2m}\big( \mathbf{P} - m (\bvarpi + \bpartial \vartheta) \big)^2 + m (U - \partial_t \vartheta) - \frac{1}{4}\, \hbar \,  \sigma(\bpartial\times\bvarpi).
\end{align*}
If $\bvarpi = - \bpartial \vartheta$, (namely, the Coriolis curvature $\Omega = d \bvarpi = i_{\bpartial\times\bvarpi}\mathrm{vol}(h)= 0$) and hence $\bpartial\times\bvarpi \equiv 0$. Remember that $\delta \Omega = \delta d \omega = \bpartial \times (\bpartial\times\bvarpi) \!\cdot\! d\bx = (\bpartial (\bpartial\!\cdot\!\bvarpi) - \Delta_{\mathbb{R}^3} \bvarpi)\!\cdot\! d\bx$.
Moreover, the self-gravitating coupling is at least modified, or if moreover $U - \partial_t \vartheta=0$ then the Newton potential can be turned off allowing the recovering the free LL equation. This makes contact with the general discussion given at the end of section~\ref{ss_barg}. In particular, the meaning of the gauge transformation on the bispinor $\varphi$ correponds to a translation $s \mapsto s + \vartheta (\bx,t)$ in the $s$ variable in the Bargmann space.

\medskip
In the usual case where we have $\bvarpi = 0$, we recover the original L\'evy-Leblond equations \cite{LL} with a scalar potential $U$, \eqref{lln} which forms, with the Poisson equation \eqref{poisson}, the L\'evy-Leblond--Newton system projected onto Newton--Cartan spacetime.

In this case, the relation between the two bispinors \eqref{rel_two_bispinors} becomes,
\begin{equation}
\label{rel_two_bispinors_varpi0}
\chi = -\frac{\hbar}{2m} \sigma(\bpartial) \, \varphi,
\end{equation}
and on replacing $\chi$ in \eqref{lln} by \eqref{rel_two_bispinors}, we recover the usual Schr\"odinger equation, for the bispinor~$\varphi$,
\begin{equation}
-\frac{\hbar^2}{2m} \Delta \varphi + m \, U \varphi = i\hbar \, \partial_t \varphi.
\end{equation}

\smallskip

%%%%%%%%%%%%%%%%%%%%%%%%%%%%%%%%%%%%%%%%%%%%%%%%%%%%%%%%%%%%%%%%%%%%
\subsection{Current and chirality}

Let us first investigate the current associated to the LLN equation.
Recall that the Bargmann structure is a relativistic structure, and for this reason, we can write the Dirac equation, although for the massless case here. We can thus define a Dirac current
\begin{equation}
j^\mu = \overline \Psi \gamma^\mu \Psi
\end{equation}
that is naturally conserved, \ie $\nabla_\mu j^\mu = 0$. What we want though, is a current on the Newton--Cartan non-relativistic spacetime. First, note that $j^0 = j^\mu \xi_\mu$, in the spatially flat case, is coherent with the definition of the mass density in \eqref{density}. Then, since $\xi$ is covariantly constant by definition, and taking into account the equivariance relation \eqref{equivariance}, we have $\nabla_s j^s = 0$. This current $(j^\mu)$ on Bargmann thus projects onto a current $(J^\alpha)$ on Newton--Cartan, which is again conserved, $\nabla_\alpha J^\alpha = 0$, with components\footnote{$\varrho$ must not be confused with $\rho = m \varphi^\dagger \varphi = m \varrho$ introduced in \eqref{poisson}.}
\begin{equation}
\varrho = \varphi^\dagger \varphi \qquad \& \qquad \boldsymbol{J} = i \left(\varphi^\dagger \bsigma \chi - \chi^\dagger \bsigma \varphi \right) \in \R^3\, ;
\end{equation}
an alternative expression of $\boldsymbol{J}$, only in terms of the principal bispinor $\varphi$, reads
\begin{equation}
\boldsymbol{J} = \frac{\hbar}{2mi} \left[ \varphi^\dagger \left(\bpartial \varphi\right) - \left(\bpartial \varphi\right)^\dagger \varphi \right] + \frac{\hbar}{2m} \bpartial \times \left(\varphi^\dagger \bsigma \varphi \right) - i \bvarpi \times \left(\varphi^\dagger \bsigma \varphi\right).
\end{equation}
We clearly notice that the first part of this current has the same general expression as the usual Schr\"odinger current, and the second part accounts for the spinorial aspect.

\medskip
Let us now turn to the study of the chirality by considering
the chiral operator $\Gamma$ acting on spinors on Bargmann space of $N=3+2$ dimensions. Since the Brinkmann metrics on Bargmann space are non diagonal, the general definition of the chiral operator has to be used,
\begin{equation}
\label{chiral}
\Gamma = - \frac{\sqrt{-g}}{5!} \epsilon_{\mu\nu\rho\lambda\sigma} \gamma^\mu \gamma^\nu \gamma^\rho \gamma^\lambda \gamma^\sigma
\end{equation}
(with the convention $\epsilon_{123ts}=+1$), which, in our case, simply gives
\begin{equation}
\label{chiral_flat}
\Gamma = I_4 \, .
\end{equation}
The triviality of the chirality operator comes from the odd dimension of Bargmann space (here $N=5$). Indeed, according to the Clifford algebra, in odd dimensions, $\Gamma$ commutes with all $\gamma^\mu$, and hence, by Schur's lemma, has to be a multiple of the identity. This is in accordance with \cite{Duv4} where the chiral operator does not seem to be relevant in non-relativistic dynamics within a space of spatial dimension 3.

%%%%%%%%%%%%%%%%%%%%%%%%%%%%%%%%%%%%%%%%%%%%%%%%%%%%%%%%%%%%%%%%%%%%
\section{LLN symmetries}
\label{sec-LLN-sym}

With the formulation of the LLN equations on a Lorentzian Bargmann spacetime, we are in position to investigate their symmetries, in particular, the maximal symmetry group.

%%%%%%%%%%%%%%%%%%%%%%%%%%%%%%%%%%%%%%%%%%%%%%%%%%%%%%%%%%%%%%%%%%%%
\subsection{Spacetime symmetries}

Finding the symmetries of the system of equations \eqref{lln_barg} is to find the transformations $\Phi$ such that if $\Psi$ is a solution of \eqref{lln_barg} then so is $\Phi^* \Psi$. 
In the following, while we explicitly show the dependence in $n$, we assume the physical case $n = 3$.

As a prerequisite, note the naturality relationship \cite{naturality} for the Dirac operator,
\begin{equation}
\label{nat_dirac}
\Phi^*(\dirac(\rg)) = \dirac(\Phi^*\rg)
\end{equation}
for all $\Phi \in \Diff(M, \rg)$, together with the naturality of the Ricci tensor \cite{Besse},
\begin{equation}
\label{nat_ricci}
\Phi^*(\Ric(\rg)) = \Ric(\Phi^*\rg)
\end{equation}
and of the equivariance operator,
\begin{equation}
\label{nat_lie}
\Phi^*(L_\xi) = L_{\Phi^*\xi}\ .
\end{equation}

From the massless Dirac equation \eqref{diracmnul}, for any transformation $\Phi$, we have $\Phi^* \left(\dirac(\rg) \Psi\right) = 0$. Introducing the naturality relationship \eqref{nat_dirac}, we have $\dirac(\Phi^* \rg) \Phi^* \Psi = 0$. To obtain the desired result, namely
\begin{equation}
\dirac(\rg) \Phi^* \Psi = 0,
\end{equation}
we need to restrict the transformations $\Phi$ to those preserving the Dirac operator, which are transformations preserving the metric up to a conformal factor, as seen with \eqref{inv_conf_dirac}. This means the $\Phi$s are such that
\begin{equation}
\label{conf_g}
\Phi^* \rg = \lambda \rg,
\end{equation}
for $\lambda$ a strictly positive valued function of $M$. Since we want the transformations to be expressed on the Newton-Cartan spacetime, the direction of the fiber generated by $\xi$ should also be preserved, hence the restriction,
\begin{equation}
\label{conf_xi}
\Phi^* \xi = \nu \xi,
\end{equation}
with $\nu$ another function of $M$.

If we want to preserve the Bargmann structure, $\Phi^* \xi$ needs to be compatible with the connection built from the transformed metric $\Phi^* \rg$. This gives the following conditions on $\lambda$ and $\nu$ \cite{BurdetDP},
\begin{equation}
d \lambda \wedge \theta = 0 \qquad \mathrm{\&} \qquad d\nu = 0.
\end{equation}
In practice, $\lambda$ turns out to be a positive non-vanishing function of time $\lambda(t)$, and $\nu \in \R$.

Let us now look at the Clifford algebra. From the equation \eqref{clialg}, we immediately get,
\begin{equation}
\Phi^* \gamma_\mu = \lambda^{\half} \gamma_\mu.
\end{equation}

From the equivariance equation \eqref{equivariance}, we have $\Phi^*\left(L_\xi \Psi\right) = \Phi^*\left(\frac{i \, m}{\hbar} \Psi\right)$. Or, with \eqref{nat_lie}, \eqref{conf_xi} and by definition of a Lie derivative, $\nu L_\xi \Phi^* \Psi = \frac{i}{\hbar} \left(\Phi^* m\right) \left(\Phi^* \Psi\right)$. If we impose the dilation of the mass parameter $m$ under these transformations, $\Phi^* m = \nu \, m$, we recover the equivariance equation for $\Phi^* \Psi$,
\begin{equation}
L_\xi \left(\Phi^* \Psi\right) = \frac{i}{\hbar} m \, \left(\Phi^* \Psi\right).
\end{equation}

To check the symmetries of the gravitation equation \eqref{poissonbarg}, we first need to learn how the density $\rho$ transforms in \eqref{density}. From the definition, $\Phi^* \rho = \Phi^*\left(m \overline\Psi^\sharp \gamma(\xi) \Psi^\sharp\right)$, we see with the help of \eqref{norm_psi} and the dilation of the mass in the paragraph above, that
\begin{equation}
\label{nat_prob_density}
\Phi^* \rho = \lambda^{-\frac{n}{2}} \nu \,m \, \overline{\left(\Phi^*\Psi\right)}^\sharp \gamma(\xi) \left(\Phi^* \Psi\right)^\sharp \ .
\end{equation}

Moving on to the last equation \eqref{poissonbarg}, we have, with \eqref{nat_ricci} $\Ric(\Phi^*\rg) = 4\pi G \left(\Phi^* \rho\right) \left(\Phi^* \theta\right) \otimes \left(\Phi^* \theta\right)$. The Ricci tensor is to be rescaled here with the conformal factor $\lambda(t)$. If we write $\lambda(t) = \phi'(t)$, then the conformal transformation law of the Ricci tensor can be put into the remarkable form \cite{SN},
\begin{equation}
\label{trsf_ricci}
\Ric\left(\phi' \, \rg\right) = \Ric(\rg) - \half (N-2) S(\phi)\, \theta \otimes \theta,
\end{equation}
where $\displaystyle S(\phi) = \frac{\phi'''}{\phi'} - \frac{3}{2}\left(\frac{\phi''}{\phi'}\right)^2 = \big(\ln\lambda \big)'' - \frac{1}{2} \big( (\ln\lambda)'\big)^2$, is the well-known Schwarzian derivative.

Upon combining the transformation law \eqref{trsf_ricci} together with the transformation of the probability density \eqref{nat_prob_density} and since $\theta = \rg(\xi)$, we obtain,
\begin{equation}
\Ric(\rg) = 4 \pi G \,m \, \nu^3 \lambda^{2-\frac{n}{2}} \, \overline{\left(\Phi^*\Psi\right)}^\sharp \gamma(\xi) \left(\Phi^* \Psi\right)^\sharp + \half (N-2) S(\phi)\, \theta \otimes \theta\ .
\end{equation}
Hence, the gravitation equation is preserved for $\Phi^* \Psi$ as long as
\begin{equation}
\label{contrainte}
\lambda^{2-\frac{n}{2}} \nu^3 = 1,
\end{equation}
(hence, $\lambda$ and $\nu$ are constant functions) and
\begin{equation}
\label{schwarzien_nul}
S(\phi) = 0.
\end{equation}
As detailed in \cite[§ 4.4 and \textit{ff.}]{SN}, this constraint which characterizes homographic transformations in time, reduces to affine time transformation as given below in \eqref{action_sn}.

\smallskip
At the end, we find that the transformations preserving the LLN system, are
\begin{equation}
\label{def_group_sym}
\mathrm{LLN}(M,\rg,\xi) = \{\Phi \in \Diff(M) \vert \Phi^*\rg = \lambda \rg, \Phi^* \xi = \nu \xi, \lambda^{2-\frac{n}{2}} \nu^3 = 1\}.
\end{equation}
The symmetrygroup of the L\'evy-Leblond--Newton equation turns out to be isomorphic to the symmetry group of the Schr\"odinger--Newton equation \cite{SN}. Thus, its action on the coordinates is given by \cite[§ 5.4.2]{SN}, for $n = 3$,
\begin{subequations}
\label{action_sn}
\begin{empheq}[left=\empheqlbrace]{align}
& \widehat{\bx} = \frac{A \bx + \bb t + \bc}{g} \\
& \widehat{t} = \frac{d t + e}{g} \\
& \widehat{s} = \frac{1}{\nu}\left( s - \langle \bb, A \bx \rangle - \frac{\Vert\bb\Vert^2}{2} t + h\right),
\end{empheq}
\end{subequations}
with $A \in \SO(3), \bb, \bc \in \R^3, d, e, g, h\in \R$, and $d \, g = \nu$.

\smallskip
Infinitesimally, this corresponds to the Lie algebra of vector fields $X$ which can be written as,
\begin{equation}
\label{conf_sn}
\left(X^\mu\right) = 
\left(
\begin{array}{l}
\displaystyle \omega \bx + t \bbeta + \bgamma + \frac{3}{n-4} \delta \bx \\
\displaystyle \frac{n+2}{n-4} \delta t + \epsilon \\
\displaystyle - \bbeta \cdot \bx - \delta s + \eta 
\end{array}
\right)
\end{equation}
with $\omega \in \so(n), \bbeta, \bgamma \in \R^n, \epsilon, \delta, \eta \in \R$ which are, respectively, generators of rotations, boosts, spatial translations, time translations, dilations, and ``vertical'' translations. For the case $n = 3$, we have $\omega \bx = \epsilon_{ijk} \omega^i x^j \boldsymbol{e}^k \equiv j(\bomega) \bx$, where $j(\bomega)$ is a skew-symmmetric matrix parametrized by $\bomega$.

\subsection{Infinitesimal actions of the LLN group}

We want to find the representation of the group action \eqref{action_sn} acting on the spinors which are solutions of the LLN equation. To this end, we will first compute the action of a Lie derivative acting on a spinor along the vector field \eqref{conf_sn} generating the Lie algebra.

To define the effect of the group action \eqref{action_sn} on objects of interests such as the gravitational potential $U$ and the Coriolis vector potential $\bvarpi$, remember that these transformations act conformally on the metric \eqref{conf_g}. We want $\wh \rg = \Phi^* \rg = \lambda \, \rg$, and since $U$ and $\bvarpi$ appear in the metric, we readily find the transformation laws \cite{SN},
\begin{equation}
\label{trsf_u_w}
\wh U(\wh \bx, \wh t) = \lambda^{-1} \nu^{-2} \left(U(\bx, t) + \bvarpi(\bx,t) \cdot A^{-1}\bb\right) \quad \mathrm{\&} \quad \wh \bvarpi(\wh \bx, \wh t) = \lambda^{-\half} \nu^{-1} \bvarpi(\bx, t) \cdot A^{-1}.
\end{equation}

Infinitesimally, the conformal condition is written as $L_X \rg = \frac{2}{N} \left(\nabla_\mu X^\mu\right) \rg$, with $X$ the vector field as in \eqref{conf_sn}. Using the general expression for the metric $\rg = \rg_0 - 2 U(\bx,t) \, dt\otimes dt + \varpi_i(\bx,t) \,dx^i\otimes dt + \varpi_i(\bx,t) \, dt\otimes dx^i$ with $\rg_0$ the flat Bargmann metric as in \eqref{barg_flat}, we obtain the Lie derivative acting on $U$ and the $\varpi_i$. Since $U$ and the $\varpi_i$ are functions, we obtain the useful relations,
%\begin{equation}
%\label{trsf_inf_u_w}
%X^\mu \partial_\mu U = -2\, \tfrac{n-1}{n-4}\, \delta \, U + \varpi_i \beta^i \quad \mathrm{\&} \quad X^\mu \partial_\mu \varpi_i = - \tfrac{n-1}{n-4}\, \delta \, \varpi_i - \epsilon_{ijk} \, \varpi_j \, \omega_k \, .
%\end{equation}
\begin{equation}
\label{trsf_inf_u_w}
X^\mu \partial_\mu U = -2\, \tfrac{n-1}{n-4}\, \delta \, U + \bvarpi \cdot \bbeta \quad \mathrm{\&} \quad (X^\mu \partial_\mu \varpi_i)\boldsymbol{e}^i  = - \tfrac{n-1}{n-4}\, \delta \, \bvarpi + \bomega \times \bvarpi \, ,
\end{equation}
where $\boldsymbol{e}^i, i=1,2,3$ is the canonical basis of $\mathbb{R}^3$.
We are now ready to compute the action of a Lie derivative of a spinor density along a conformal vector field $X$. Developing the expression of a Lie derivative of a spinor density \eqref{lie_spinor_density} in terms of partial derivatives, we get
\begin{equation}
\begin{split}
L_X \Psi = \, & X^\mu \partial_\mu \Psi - \frac{1}{8} X^\mu \left[\gamma^\rho, \partial_\mu \gamma_\rho\right] \Psi + \frac{1}{8} X^\mu \left[\gamma^\rho, \Gamma^\sigma_{\mu\rho} \gamma_\sigma\right] \Psi - \frac{1}{8} \left[\gamma^\mu, \gamma^\nu\right] \partial_\mu X_\nu \Psi + \\
&  + \frac{N-1}{2N} \, \partial_\mu X^\mu \Psi + \, \frac{N-1}{2N} \Gamma^\mu_{\mu\lambda} X^\lambda \Psi
\end{split}
\end{equation}
for any conformal Killing vector field $X$.

Computing all these terms for the expression of the vector field \eqref{conf_sn}, for $n = 3$, and in view of \eqref{contrainte} and \eqref{trsf_inf_u_w} we find the expression,
\begin{equation}
\label{inf_rep}
L_X \Psi = \underbrace{X^\mu \partial_\mu}_\textrm{\encircle{1}} \Psi + 
\underbrace{\left(
\begin{array}{cc}
-\frac{n-1}{2(n-4)} \, \delta + \frac{i}{2} \sigma(\bomega) & 0 \\[2ex]
\frac{i}{2} \sigma\left(\bbeta\right) & \frac{n-1}{2(n-4)} \, \delta + \frac{i}{2} \sigma(\bomega)
\end{array}
\right)}_\textrm{\encircle{2}}  \Psi
+ \underbrace{\frac{3(n+1)}{2(n-4)} \, \delta}_\textrm{\encircle{3}} \, \Psi \ .
\end{equation}

\smallskip
These conformal transformations thus act in three parts on our spinors:
\begin{enumerate}[1)]
\item
The first part is the coordinate transformation, \ie $\Psi(x, t, s) \rightarrow \Psi(\wh x, \wh t, \wh s)$.
\item
The second part of the transformation is the mixing of the two bispinors under rotations, and the fact that the two bispinors are dilated separately under these transformations. 
\item
The last part of the transformation comes from the dilation of the volume of the densities. This is a global factor encompassing the two bispinors.
\end{enumerate}

%%%%%%%%%%%%%%%%%%%%%%%%%%%%%%%%%%%%%%%%%%%%%%%%%%%%%%%%%%%%%%%%%%%%
\subsection{Integration to group representation}

To obtain a representation of the LLN group through \eqref{inf_rep}, is to find $\rho(\Phi) \Psi = (\Phi^{-1})^* \Psi$, such that if $\Psi$ is a solution of the LLN system \eqref{lln_barg}, then $\rho(\Phi) \Psi$ is again a solution.

\smallskip
The first step is thus to find the reverse action of \eqref{action_sn}, \ie $(\wh \bx, \wh t, \wh s) = \Phi^{-1}(\bx, t, s)$, for $\Phi = (a, \bb, \bc, d, e, g, h)$ belonging to the LLN group, where $a \in \mathrm{SU}(2)$ is such that $a \, \sigma( \bx) \, a^{-1} = \sigma  (A \bx)$. We get \cite{SN},
\begin{subequations}
\label{rev_action_sn}
\begin{empheq}[left=\empheqlbrace]{align}
& \wh \bx = A^{-1} \left[g \bx - \frac{g t - e}{d} \bb - \bc\right] \\[2mm]
& \wh t = \frac{gt - e}{d} \\[2mm]
& \wh s = \nu s + g \left\langle\bb, \bx\right\rangle - \frac{g}{2d} \Vert\bb\Vert^2 t + \frac{e}{2d} \Vert\bb\Vert^2 - \left\langle \bb, \bc \right\rangle - h
\end{empheq}
\end{subequations}

\smallskip\noindent
with $d = \nu^{\frac{n-1}{n-4}}$ and $g = \nu^{-\frac{3}{n-4}}$.

\smallskip
First, if we restrict ourselves to the subgroup of dilations, we have, using \eqref{inf_rep} and \eqref{decomp_psi},

\begin{equation}
\label{rep_dil2}
\left[\rho(u_\nu) \psi\right](\bx, t) = \nu^{-\frac{3(n+1)}{2(n-4)}} \, \left(
\begin{array}{cc}
\nu^{\frac{n-1}{2(n-4)}} & 0 \\
0 & \nu^{-\frac{n-1}{2(n-4)}}
\end{array} \right) \,
\psi\left(\nu^{-\frac{3}{n-4}} \bx, \nu^{-\frac{n+2}{n-4}} t\right)
\end{equation}
where we find again the three elements of the conformal transformations. From left to right: the global factor coming from the dilation of the volume element; then the matrix transforming the two bispinors, which can also be put in the remarkable form $\left(\begin{array}{cc}d^\half & 0 \\ 0 & d^{-\half}\end{array}\right)$; then the action on the coordinate variables. Hence the dynamical exponent of this model,
\begin{equation}
\label{dyn_exp}
z = \frac{N}{3} = \frac{n+2}{3},
\end{equation}
which is the same as in the Schr\"odinger--Newton case as found in \cite{SN}. This ought to be expected as we can recover the same form of the (generalized) Schr\"odinger--Newton equation \eqref{eq-H}, though for a bispinor and with a spin contribution.

In the case of $n = 3$, we get the representation $\left[\rho(u_\nu) \psi\right](\bx, t)~=~\nu^{6} \left(\begin{array}{cc} \nu^{-1} & 0 \\ 0 & \nu\end{array}\right) \psi(\nu^3 \bx, \nu^5 t)$, and thus $z = 5/3$.

Let us now consider a general element of the LLN group of the form $u(a, \bb, \bc, d, e, g, h)$. We can extract the dilations, acting with $d$ and $g$, using the decomposition,
\begin{equation}
\label{decomposition_u}
u(a, d^{-1} \bb, g^{-1} \bc, 1, g^{-1} e, 1, (dg)^{-1} h) \cdot u_\nu(1, 0, 0, d, 0, g, 0) = u(a, \bb, \bc, d, e, g, h).
\end{equation}

The left element above, without dilations, belongs to the Bargmann subgroup, which is the group of isometries of a Bargmann structure $(M, \rg, \xi)$. For such element of the form $u_B(a, \bb, \bc, 1, e, 1, h)$, we have the known representation \cite{LL},
\begin{equation}
\label{rep_barg}
\left[\rho(u_B) \psi\right](\bx, t) = \exp\left( \frac{im}{\hbar} \left(\angles{\bb, \bx-\bc} - \frac{\Vert\bb\Vert^2 }{2}( t-e) - h\right) \right)
\left(
\begin{array}{cc}
a & 0 \\
-\frac{i}{2}  \sigma (\bb) \, a & a
\end{array} \right)
\psi\left(\wh \bx, \wh t\right).
\end{equation}
It is worthwhile to notice that the transformation (\ref{rev_action_sn}c) yields the phase factor.

When combining the two representations \eqref{rep_barg} and \eqref{rep_dil2} by using the decomposition \eqref{decomposition_u} we then get for the full action of the LLN group on bispinor. For a general element $u(a, \bb, \bc, d, e, g, h)$ of the LLN group, one has the following projective unitary representation
\begin{equation}
\label{full_rep}
%\fbox{$\displaystyle
\boxed{
\begin{aligned}
\left[\rho(u) \, \psi\right](\bx, t) = & \, \nu^{-\frac{3(n+1)}{2(n-4)}} \, \exp\left( \frac{im}{\nu \hbar} \left(g \angles{\bb, \bx} - \frac{g}{2d} \Vert\bb\Vert^2 t + \frac{e}{2d} \Vert\bb\Vert^2 - \angles{\bb, \bc} - h\right) \right)\\[2mm] 
& \left(
\begin{array}{cc}
\nu^{\frac{n-1}{2(n-4)}} \, a & 0 \\
-\frac{i}{2} \sigma (d^{-1} \bb) \, a & \nu^{-\frac{n-1}{2(n-4)}} \, a
\end{array} \right) \,
\psi\left(gA^{-1} \bx - \frac{gt-e}{d}\, \bb - \bc \, , \, \frac{gt-e}{d}\right)
\end{aligned}
}%$}
\end{equation}
once again with $d = \nu^{\frac{n-1}{n-4}}$ and $g = \nu^{-\frac{3}{n-4}}$ (with $n=3$).

It can be verified that $\rho(u) \psi$ is indeed a solution of the generalized LLN equation \eqref{generalized_lln_eq} if $\psi$
is.\footnote{\samepage This can be seen at the infinitesimal level with the Lie derivative $L_X$ on spinor densities \eqref{lie_spinor_density} along a conformal Killing vector field $X$, \ie such that $\nabla_{(\mu} X_{\nu)} = \frac{1}{N} (\nabla_\rho X^\rho) \rg_{\mu\nu}$, and of the Dirac operator $\dirac(\rg)$ \eqref{dirac_spinor} on a spinor density, we find the commutator,
\begin{equation*}
\label{lxdirac}
\left[ L_X, \dirac(\rg)\right] \Psi = \frac{N-1}{2N} (\nabla_\mu X^\mu) \, \dirac(\rg) \Psi.
\end{equation*}
This means that whenever $\Psi$ is a solution of $\dirac(\rg) \Psi = 0$, then so is $\Psi_\epsilon \equiv \Psi + \epsilon L_X \Psi + \ldots$, for any conformal Killing vector field $X$.
}
This is also true for the LLN equation \eqref{lln} without the Coriolis vector potential.

%%%%%%%%%%%%%%%%%%%%%%%%%%%%%%%%%%%%%%%%%%%%%%%%%%%%%%%%%%%%%%%%%%%%
\subsection{Action, energy-momentum tensor and conserved quantities}

To obtain the symmetries of the system, one way to proceed is through an action principle. Having succeeded in adapting the LLN system \eqref{lln_barg} to a Bargmann structure, it is natural to define the action principle on the Bargmann manifold $M$. Since the wave equation \eqref{diracmnul} is what determines the time evolution of the system, we will consider its action $S_D$, while the other equations, notably the gravitational equation and the equivariance relation are postulated without deriving them from an action principle. A justification for this could be that both the gravitational equation and the equivariance are inherent to the Bargmann structure, in the sense that they stem from its geometry.

\smallskip
Thus, for the massless Dirac equation \eqref{diracmnul}, we have the usual action of the Dirac equation in curved spacetime, with here zero mass,
\begin{equation}
\label{action_dirac}
S_D[\psi,\rg] = i \hbar \int_M \overline \psi \diracf \psi \volg d^Nx.
\end{equation}

Variations of the fields lead to the massless Dirac equation, and variations of the metric gives us the energy-momentum tensor (EMT), which we will use to compute conserved quantities. Recall its definition,
\begin{equation}
\label{def_emt}
T_{\mu\nu} = - \frac{2}{\volg} \frac{\delta S_D}{\delta \rg^{\mu\nu}}.
\end{equation}
The EMT obtained from \eqref{action_dirac} is much simpler than the one for SN. After symmetrization, it is given by
\begin{equation}
\label{emt}
T_{\mu\nu} = \frac{i \hbar}{4} \left(\overline \psi \gamma_\mu \nabla_\nu \psi + \overline \psi \gamma_\nu \nabla_\mu \psi - \nabla_\mu \overline \psi \gamma_ \nu \psi - \nabla_\nu \overline \psi \gamma_\mu \psi\right).
\end{equation}
This expression of the EMT for spinors already appears in \cite{Weldon:2000fr}.

\medskip
The next step is now to compute the conserved currents and quantities associated to the EMT (\ref{emt}) and the conformal symmetries \eqref{conf_sn}. To build these, a method similar to Souriau's is used~\cite{Souriau74}. See also \cite{Duval:1994pw,Duval:1975dj,DHP}.

Diffeomorphisms act infinitesimally on the Lagrangian $\cL_D$, defined as $S_D = \int_M \cL_D \volg d^Nx$, associated to the action functional \eqref{action_dirac}, by
\begin{equation}
L_X \left(\cL_D \volg\right) = (\nabla_\mu X^\mu) \cL_D \volg + \left(L_X \cL_D\right) \volg\, ,
\end{equation}
with $X \in \Vect(M)$.

On the equations of motion, we have $\cL_D = 0$. Then, since $\cL_D$ is represented by a closed N-form, we have by Cartan's formula that $L_X \cL_D = d\left(i_X \cL_D\right)$. Hence, on the equations of motion, $L_X \left(\cL_D \volg\right) = d\left(i_X \cL_D\right) \volg$. Thus,
\begin{equation}
L_X S_D = 0\, .
\end{equation}

With an action invariant under diffeomorphisms, the EMT is automatically divergence free. Indeed, from the definition of the EMT \eqref{def_emt}, we have, $0 = L_X S_D = \half \int_M T^{\mu\nu} (L_X \rg)_{\mu\nu} \volg d^Nx$. From the definition of a Lie derivative, the EMT being symmetric, and an integration by parts, we have $0 = \int_M \left(\nabla_\mu T^{\mu\nu}\right) X_\nu \volg d^Nx, \, \forall X \in \Vect(M)$. Hence the well known result for the EMT of the Dirac equations,
\begin{equation}
\nabla_\mu T^{\mu\nu} = 0.
\end{equation}
This can also be computed directly with the help of the field equation, and the various symmetries of the Riemann tensor. Also, through the field equations, we clearly have that the energy-momentum tensor is traceless, or $\rg^{\mu\nu} T_{\mu\nu} = 0$.

We now have all the ingredients to build up conserved charges. We want to build currents $k = (k^\mu)$ that are conserved, \ie $\nabla_\mu k^\mu = 0$. Two objects are of particular interest here: the EMT \eqref{emt}, which is divergence-free and traceless, and the conformal Killing vector field $X^\nu$ associated to the conformal symmetries of our system. Now, a current built as:
\begin{equation}
\label{k_barg}
k^\mu = T^{\mu\nu} X_\nu
\end{equation}
is conserved. Indeed, by taking the divergence of this expression, and using the fact that $\nabla_{(\mu} X_{\nu)} = \cL_X \rg_{\mu\nu} = \lambda \rg_{\mu\nu}$ for a conformal Killing field, and the properties that the EMT is traceless, symmetric, and divergent free, we have,
\begin{equation}
\label{cons_curr}
\nabla_\mu k^\mu = 0.
\end{equation}

However, for now, $k^\mu$ lives in Bargmann space, of dimension $N = n + 2$, but we would like conserved currents on the non relativistic spacetime. Notice that the action does not depend on $s$, it is $\xi$-invariant. The same goes for the EMT, but unlike \cite{DHP}, here we have $\nabla_s X^s \neq 0$ , because of the dilations. Thus, the current $k^\mu$ does not project onto spacetime here. However, to get a charge living on NC spacetime, we can integrate the current on $\widetilde{\Sigma}_t$, \ie on both space and the fiber of Bargmann spacetime, instead of only space $\Sigma_t$.

Since $k^0 = \xi_\mu k^\mu$, the conserved charges read as:
\begin{equation}
\label{cons_charges}
Q_X = \frac{1}{2 \pi} \int_{\widetilde{\Sigma}_t} T_{\mu\nu} X^\nu \xi^\mu \sqrt{\rg_{{}_{\Sigma_t}}}\, \mu(s) \, d^n x \, ds,
\end{equation}
with $\mu(s)$ the integration measure of the variable $s$. Indeed, this is a time like dimension, and we can choose the fiber to be $S^1$ instead of $\R$, so that $\widetilde{\Sigma}_t = \Sigma_t \times S^1$, to get convergent integrals. If $\theta \in (-\pi,+\pi)$ is the angular coordinate on $S^1$, then $s = 2 \tan(\theta/2)$ is an affine coordinate. The integration measure is thus $\mu(s) = \frac{1}{1+s^2/4}$. Note that $\int^\infty_{-\infty} \mu(s) ds = 2 \pi$, and $\int^\infty_{-\infty} s \mu(s) ds = 0$. Most charges do not depend on $s$, and thus only get a $2 \pi$ factor. The only exception is for the charge associated to dilations, for $n \neq 4$, in which case the contribution linear in $s$ in the integrand will disappear after integration.

Altogether, this is the formulation of the Noether theorem applied to Bargmann structures.

\smallskip
Since there is one conserved quantity for each generator of the Lie algebra of the LLN group, one can write $Q_X$ as
\begin{equation}
Q_X = \boldsymbol{J}\cdot\boldsymbol{\omega}  + \boldsymbol{P} \cdot \bgamma + \boldsymbol{G} \cdot \bbeta + H \, \epsilon + D \, \chi + M \, \eta.
\end{equation}

Computing \eqref{cons_charges}, we find the following conserved charges, for $n = 3$ and in the flat case, but with Coriolis forces,

\begin{equation}
\label{conserved_n_neq4}
\left\lbrace
\begin{array}{ll}
\displaystyle E = \int \varphi^\dagger H \varphi \, d^3 \bx \qquad \qquad & \mathrm{energy} \\[4mm]
\displaystyle \boldsymbol{P} \equiv \int \boldsymbol{\cP} \, d^3 \bx = \frac{i\hbar}{2} \int \left((\bnabla \varphi)^\dagger \varphi - \varphi^\dagger \bnabla \varphi - i \, m \, \bvarpi\times (\varphi^\dagger \boldsymbol{\sigma} \varphi)\right) d^3 \bx & \mathrm{linear\; momentum} \\[4mm]
\displaystyle \boldsymbol{J} = \int \bx \times \boldsymbol{\cP} \, d^3 \bx + \frac{\hbar}{2} \int \varphi^\dagger \boldsymbol{\sigma} \varphi \, d^3\bx & \mathrm{angular\; momentum} \\[4mm]
\displaystyle M = m \int \varphi^\dagger \varphi \, d^3 \bx & \mathrm{mass} \\[4mm]
\displaystyle \boldsymbol{G} = t \boldsymbol{P} - m \int \varphi^\dagger \varphi \, \bx \, d^3 \bx & \mathrm{boost} \\[4mm]
\displaystyle D = \frac{n + 2}{n-4}\, t E + \frac{3}{n-4} \int \bx \cdot\! \boldsymbol{\cP} \, d^3 \bx & \mathrm{dilation}\ (n=3)
\end{array}
\right.
\end{equation}
with $H$ the Hamiltonian given in \eqref{hamiltonien}. Notice that in the conserved quantity $D$ with $n=3$, the dynamical exponent $z=5/3$ is split into $-5$ for the time part and $-3$ for the space part.

\smallskip\noindent
These conserved quantities are qualitatively the same as for Schr\"odinger--Newton \cite{Marsot}, with two slight differences. We now have a bispinor $\varphi$ instead of a scalar wave-function, and we have a new contribution to the angular momentum due to the spin.
Here, once again, we note that the second bispinor plays no role, only the first one, $\varphi$, is important. These conserved quantities must also be compared with those obtained in~\cite{Duval:1994pw}.

%%%%%%%%%%%%%%%%%%%%%%%%%%%%%%%%%%%%%%%%%%%%%%%%%%%%%%%%%%%%%%%%%%%%
\section{Conclusion}
\label{sec-conclusion}

In order to facilitate the study of the symmetries of the L\'evy-Leblond--Newton equations within a geometric view, we recast the later on Bargmann structures which are Lorentzian manifolds of one dimension higher than the studied Newton-Cartan space-time. This allows us to write the LLN equations in a completely covariant formulation, notably involving a massless Dirac equation on the Bargmann structures. In addition, this geometrical framework yields a natural generalization of the L\'evy-Leblond--Newton equations, where Coriolis forces can be taken into account. Despite the self-coupling of the spinor with itself by gravity, and the Coriolis forces, the second bispinor remains non-dynamical, in accordance with L\'evy-Leblond's remarks~\cite{LL}. This is to be expected %considering
since the Schr\"odinger equation is first order in time and the LL equation is morally its ``square root''. To some extent, the physical interpretation of this second bispinor in the non-relativistic framework deserves to be better understood.

Thanks to the geometrical framework of Bargmann structures and the covariant rewriting of the LLN equations, we were able to find the maximal symmetry group of this system which turns out to be the same as that of the Schr\"odinger--Newton equations, namely the SN group~\cite{SN}. This group is of dimension 12 in 3+1 dimensional space-time. The action of this group on 4-component spinors was computed, and of particular interest is the scaling law of the theory: in 3+1 dimensions, the dynamical exponent turns out to be $z = 5/3$. This is the same unusual dynamical exponent as in the Schr\"odinger--Newton case which also occurs in \cite{darkon}. It is a curiosity that the dynamical exponent $z=N/3$ obtained in \eqref{dyn_exp} (with $n\neq 4$) conserves a trace of the $N$-dimensional Bargmann space. Finally, we computed the conserved quantities associated to the symmetries of the generalized LLN system driven by the SN group. They depend on the main dynamical bispinor.

\smallskip
As a final comment, both the SN equation and the LLN equation, tackled within the same geometrical approach, have the SN group as maximal symmetry group and yield $z = 5/3$ for the 5-dimensional Bargmann space.
More generally, there exists a notion of \textit{generalized symmetry} of the Schr\"odinger equation associated with the infinite-dimensional \textit{Schr\"odinger-Virasoro} group \cite{Hen,RU}. This symmetry has been fully geometrized in~\cite{SV} in which the automorphisms of conformal Bargmann structures are investigated. This provides a relationship between the Schr\"odinger--Virasoro group and the extended symmetries of the Schr\"odinger--Newton group which apply to both the SN equation and the LLN equation as specific examples.

%%%%%%%%%%%%%%%%%%%%%%%%%%%%%%%%%%
\section*{Acknowledgments}
%%%%%%%%%%%%%%%%%%%%%%%%%%%%%%%%%%

This work has widely benefited from the expertise of our late colleague and collaborator, C.~Duval, whose inputs were so valuable. We are also indebted to J.-Ph.~Michel for discussion and suggestions about the generalized LLN equation.

The project leading to this work has received funding from ExcellenceInitiative of Aix-Marseille University - A*MIDEX, a French ``Investissements d’Avenir'' programme. %%%%%%%%%%%%%%%%%%%%%%%%%%%%%%%%%%%%%%%%%%%%%%%%%%%%%%%%%%%%%%%%%%%%%%%%%%%%
%%%%%%%%%%%%%%%%%%%%%%%%%%%%%%%%%%%%%%%%%%%%%%%%%%%%%%%%%%%%%%%%%%%%%%%%%%%%
%\bibliographystyle{unsrt}
%\bibliographystyle{alpha}
\bibliographystyle{plain}

\newcommand\doi[1]{\href{http://dx.doi.org/#1}{DOI: #1}}

%%%%%%%%%%%%%%%%%%%%%%%%%%%%%%%%%%%%%%%%%%%%%%%%%%%%%%%%%%%%%%%%%%%%%%%%%%%%
%%%%%%%%%%%%%%%%%%%%%%%%%%%%%%%%%%%%%%%%%%%%%%%%%%%%%%%%%%%%%%%%%%%%%%%%%%%%

\end{document}